\documentclass[10pt,conference]{IEEEtran}

\IEEEoverridecommandlockouts
\usepackage[cmex10]{amsmath}
\usepackage{stfloats}
\usepackage{amssymb}
\usepackage{stackrel}

\usepackage{url}

\usepackage{graphicx}
\usepackage{cite}
\usepackage{enumerate}
\usepackage{url}

\newtheorem{theorem}{Theorem}
\newtheorem{lemma}{Lemma}
\newtheorem{remark}{Remark}

\newcommand{\Rmnum}[1]{\expandafter\@slowromancap\romannumeral #1@}

\begin{document}

\title{Outage Performance in Secure Cooperative NOMA}

\author
{
Milad Abolpour, Mahtab Mirmohseni and Mohammad Reza Aref\\
Information Systems and Security Lab (ISSL) \\
Department of Electrical Engineering, Sharif University of Technology, Tehran, Iran\\
Email: miladabolpour@ee.sharif.edu,\{mirmohseni,aref\}@sharif.edu\thanks{This work was partially supported by Iran National Science Foundation
(INSF) under contract No. 96/53979.
}}

\maketitle

\begin{abstract}
Enabling cooperation in a NOMA system is a promising approach to improve its  performance. In this paper, we study the cooperation  in a secure NOMA system, where the legitimate users are distributed uniformly in the network and the eavesdroppers are distributed according to a homogeneous Poisson point process. We consider  a  cooperative NOMA scheme (two users are paired as strong and weak users) in two phases: 1) \nolinebreak Direct transmission phase, in which the base station broadcasts a superposition of the messages, 2) Cooperation phase, in which the strong user acts as a relay to  help in forwarding the messages of the weak user.
We study the secrecy outage performance   in two cases: (\textit{i}) security of the strong user, (\textit{ii}) security of  both users,
 are guaranteed. In the first case, we derive the exact secrecy outage probability of the system for some regions of power allocation coefficients and a  lower bound on the secrecy outage probability is derived for the other regions. In the second case, the strong user is a relay or a friendly jammer (as well as a relay), where an upper bound on the secrecy outage probability is derived at high signal-to-noise-ratio regimes.  For both cases, the cooperation in a two-user paired  NOMA system necessitate to utilize the joint distribution of the distance between two random users. Numerical results shows the superiority of the secure cooperative NOMA for a range of the cooperation power compared to secure non-cooperative NOMA systems.

\end{abstract}


\section{Introduction}
\IEEEPARstart{R}ECENTLY, non-orthogonal multiple access (NOMA)  systems have been popular in fifth-generation (5G) networks due to their high power spectral efficiency. In NOMA systems, users are classified into orthogonal multiple access (OMA) groups. In each group, by accommodating several users within the same resource blocks, such as frequency and time,  the significant bandwidth and also the latency of users are decreased.  Base station (BS) superimposes the messages and  the stronger user exploits successive interference cancelation (SIC) \cite{islam2017power}. Ding \textit{et al}. in \cite{ding2014performance}, investigated the performance of a NOMA system with random deployed users.  In this scheme, outage probability (OP) was used to demonstrate that under a condition on power allocation coefficients and users' targeted data rates, NOMA can achieve a diversity order as an  orthogonal multiple access  system and in the case of ergodic sum rates, NOMA has a better performance than OMA systems.

In NOMA systems, existence of weak users degrades the outage performance of the system. Exploiting device-to-device (D2D) transmission capability of 5G users, enabling cooperative NOMA, enhances the efficiency of NOMA systems. Effects of the cooperative transmission in NOMA, by applying outage probability  as a metric, was investigated by Ding \textit{et al.} \cite{ding2015cooperative}, where demonstrated that due to the complexity limitations of the system, utilizing all the users in  a cooperative NOMA system is not an efficient way but pairing users with more distinctive channel coefficients provides higher gain.

In wireless networks, signals are transmitted in open to all network users, so the security of the users must be provided. Using the physical layer capabilities is a promising way to maintain the security of the users. Signals are overheard by external or internal eavesdroppers. Secrecy performance of a cellular NOMA network in cases of single and multiple-antenna BS was investigated in \cite{liu2017enhancing}, where a random number of external passive eavesdroppers are distributed according to a  homogeneous Poisson point process (PPP) across a circular area. Lei \textit{et al.} investigated the secrecy outage probability (SOP) of a NOMA system containing two users, multiple-antenna BS and an external  eavesdropper, who  overhears only one of the users \cite{lei2017secure}.
The case of internal eavesdroppers in a cooperative NOMA system is studied in \cite{bassem2018untrusted}.

Although cooperation in NOMA systems enhances the outage performance, it gives more opportunity to the eavesdroppers to overhear the messages of the weaker users in the network. Chen \textit{et al.} studied the secrecy of a cooperative NOMA system with a  Decode-and-Forward  (DF) and an Amplify-and-Forward (AF) relay in existence of one eavesdropper. It was shown that at high signal-to-noise -ratio (SNR) regimes, DF and AF relays have the  same performances \cite{chen2018physical}. Zheng \textit{et al.} investigated the secrecy in a network consists of two users, a relay and some eavesdroppers  \cite{zheng2018secure}, in which the relay transmits the messages and generate an artificial noise in order to decrease the SNR of the eavesdroppers to decode the messages of the legitimate users (LUs).

Though using external relay nodes in order to realize cooperation in secure NOMA systems has been studied in some works [7-8],  this cooperation is also possible by using internal nodes, where the strong user acts as a relay to help in forwarding the signal of the weak user as studied in \cite{bassem2018passive} for a simple network with a Base station, two LUs and one eavesdropper at high SNR regimes \cite{bassem2018passive}.

 In this paper, we investigate the secrecy performance of a cooperative  NOMA network with many LUs in existence of a random number of external passive eavesdroppers. We assume that every LUs are paired randomly (called strong and weak users) and we analyze the secrecy performance of one pair. We consider two cases: the security is provided (1)   for the message of the strong user, when the targeted data rate of the strong user is greater than the weak user; (2) for the  messages of both users. In case 1, we derive a lower bound on the SOP, which is tight in some regions of the power allocation coefficients and users' targeted data rates. In case 2, we derive an upper bound on the SOP of the system at high SNR regimes. In this case, we propose two strategies: in the first strategy the strong user acts a relay and allocates all of its transmitting power for sending the message of the weak user, while in the second strategy, the strong user acts as a friendly jammer (as well as a relay), where it allocates a proportion of its transmitting power to send noise and the  rest of its transmitting power is allocated to send the message of the weak user.

\begin{small}
\textbf{\textit{Notation:}}  $F_{X}\left( x\right)$ and $f_{X}\left( x\right)$ are the cumulative distribution function (CDF) and the  probability density function (PDF) of random variable $x$, respectively. $\overline{A}$  is the complementary event of the event $A$, where $\text{Pr} \left(\overline{A}\right)=1-\text{Pr} \left( A\right)$, $\Gamma \left( . \right)$ is the gamma function, where $\Gamma \left( s \right)=\int\limits_{0}^{\infty} t^{s-1}e^{-t} \mathrm{d}t$ and $\Gamma \left(. \, , \, . \right)$ is the upper incomplete gamma function, where $\Gamma \left( p,q \right)= \int\limits _{q}^{\infty} t^{p-1} e^{-t} \mathrm{d}t$.
\end{small}

\section{System Model}

We consider  $S_{1}$ as the eavesdropper-free zone with radius $r_{p}$,  $S_{1}$ and $S_{2}$ as the user zone with radius $r_{p}$ to $r_{l}$, $S_{2}$ and $S_{3}$ as the eavesdropper zone with radius $r_{l}$ to $r_{e}$,  as depicted in Fig. 1. Our system consists of a single antenna base station which is located at the center of the  $S_{1}$, $n_{l}$  LUs distributed uniformly in the  user zone  and  a random number of the eavesdroppers distributed according to   a homogeneous PPP, which is denoted by $\Phi_{e}$  with the density  $\lambda_{e}$, in the eavesdropper zone. The system model resembles the one in \cite{liu2017enhancing}.

 Channel state information (CSI) of each  LU is known at the BS, the eavesdroppers and the other LUs but CSI of the eavesdroppers are unknown at the BS and LUs.
\begin{figure}[t]
\centering
\includegraphics[width=6cm,height=6cm]{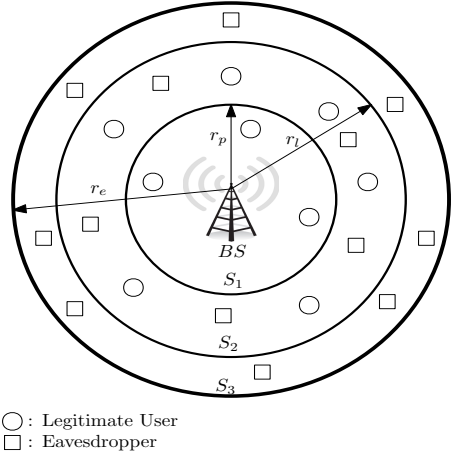}
\caption{Network model for secure cooperative NOMA transmission}
\end{figure}
All channels assumed to experience quasi-static Rayleigh fading, where the channel coefficients are constant for each transmission block but vary independently  between different blocks. LUs are ordered according to their channel coefficients as $ |h_{n_{l}}|^{2} \leq \cdots \leq |h_{2}|^{2}\leq |h_{1}|^{2}$, where $h_{i}=\frac{ g_{BS,i}}{\sqrt{ 1+d_{BS,i}^{\alpha}}}$, in which  $g_{BS,i}$ denotes the Rayleigh fading channel gain  between  user $i$ and the $BS$, $d_{BS,i}$ denotes the distance between user $i$ and the  $BS$ and also $\alpha$ is the path-loss exponent.  In this model, every two users are paired  randomly with each other and they make cooperative NOMA systems and we  investigate the secrecy performance of one pair. We remark that each resource block is devoted to two paired users based on the NOMA scheme. Transmitting the  messages of the users contains two phases: 1) Direct transmission phase, \linebreak
 2) Cooperation phase. 

Consider  a strong user, $ U_{m} $,  is paired with a weak user, $U_{n}$. We study two different cases. First, the message of  $U_{m}$ is sent securely and second, the messages of both users are sent securely.
  \subsection{Direct Transmission Phase}
In this phase, the $BS$  broadcasts a linear combination of the  messages of  $U_{m}$ and $U_{n}$  with total power of  $ P_{BS}$  as:
   \begin{align*}
   X_{BS}^{\left( 1 \right)}=\left( a_{m}S_{m}+a_{n}S_{n} \right) \sqrt{P_{BS}},
   \end{align*}
 where $a_{m} , a_{n}$ denote  the power allocation coefficients for  $U_{m}$ and $U_{n}$, respectively and $S_{m}, S_{n}$ are the messages of  $U_{m}$ and $U_{n}$, respectively. By following the NOMA protocols, we assume that $a_{n}> a_{m}$ and $  a_{n}^{2}+ a_{m}^{2}=1$.

The received signals at  LUs and the eavesdroppers are:
  \begin{equation*}
  Y_{j}^{(1)}=h_{j}\left( a_{m}S_{m}+a_{n}S_{n}\right)\sqrt{P_{BS}} +N_{j}^{(1)},
  \end{equation*}
 where $j \in \left\lbrace m ,n ,e \right\rbrace,$ $ N_{j}^{(1)}$  is a zero-mean additive white Gaussian noise (AWGN) with variance $\sigma_{j}^{2}$, $h_{e}=\frac{g_{BS,e}}{\sqrt{1+d_{BS,e}^{\alpha}}}$ denotes the  channel coefficient,  in which $g_{BS,e}$ is the fading channel gain  between the $BS$ and the eavesdroppers and $d_{BS,e}$ is the distance between the $BS$ and eavesdroppers. At the end of this phase, $U_{m}$ performs SIC and decodes its own message. The SNR of the $U_{m} \left( \gamma_{m}^{m,\left(1\right)} \right)$ and the maximum SNR of an eavesdropper $ \left(\gamma _{e}^{m,\left(1\right)} \right)$ to decode $S_{m}$ are shown in the following:
  	\begin{align*}
  	\gamma _{m}^{m,\left(1 \right)} &=a_{m}^{2}\, \frac{P_{BS}}{\sigma_{m}^{2}} \,  |h_{m}|^{2}  , \\
    \gamma _{e}^{m,\left(1\right)} &=a_{m}^{2}\, \frac{P_{BS}}{\sigma_{e}^{2}} \, |h_{e}|^{2} , \ e\in \Phi_{e}.
    \end{align*}
  \subsection{Cooperation Phase}
  	\subsection*{\textbf{Case 1:} Maintaining Secrecy at the Stronger User}
  	
  	 If $U_{m}$ is not  able to perform SIC, system stops working. When  $U_{m}$ is able to decode $S_{n}$, this phase starts and  $U_{m}$ transmits the message $S_{n}$ with the power $P_{C}$. Therefore, the  transmitted signal of $U_{m} \left(X_{m}\right) $ is shown below:
\begin{align}
\label{eq: Direct phase}
X_{m}^{\left( 2 \right)}=\sqrt{P_{C}}S_{n}.
\end{align}
The Received signal at $U_{n}$ equals to:
\begin{align*}
Y_{n}^{\left( 2 \right)}= \sqrt{P_{C}} \frac{g_{m,n}}{\sqrt{1+d_{m,n}^{\alpha}}}S_{n}+N_{n}^{\left( 2\right)},
\end{align*}	
\begin{figure*}[!b]
\hrulefill
 \setcounter{equation}{2}
\begin{equation}\label{PDFe}
\begin{aligned}
f_{\gamma _{e}^{m,\left(1 \right)}} \left( x \right) = \mu _{1}e^{-\frac{x}{P_{BS}a_{m}^{2}}} \exp \left[ -\frac{\mu_{1}e^{-\frac{x}{P_{BS}a_{m}^{2}}} \Gamma \left( \eta \, , \, \mu_{2}x   \right)}{x^{\eta}} \right]
 \left( \frac{\mu_{2}^{\eta}e^{-\mu_{2}x}}{x}+\frac{\eta  \Gamma \left( \eta \, , \, \mu_{2}x   \right) }{x^{\eta +1}} + \frac{\Gamma \left( \eta \, , \, \mu_{2}x   \right)}{P_{BS}a_{m}^{2} x^{\eta}}\right).
\end{aligned}
\end{equation}
\end{figure*}
where  $g_{m,n}$ is the channel coefficient between $U_{m}$ and $U_{n}$ with exponential distribution  with parameter $\lambda _{m,n}$, $d_{m,n}$ denotes the distance between $U_{m}$ and $U_{n}$ and $N_{n}^{\left( 2\right)}$ is a zero-mean  AWGN with variance $\sigma_{n}^2$.  $U_{n}$ uses the maximum ratio combining (MRC) receivers \cite{tse2005fundamentals} to decode $S_{n}$, therefore the signal-to-noise-plus-interference (SINR) of $U_{n} \left( \gamma_{n}^{n} \right)$ to decode $S_{n}$ equals to:
\begin{align*}
\gamma _{n}^{n}=\gamma _{n}^{n,\left( 1 \right)}+\min \left(\gamma _{n} ^{n,\left( 2 \right)}  \, ,\, \gamma _{m}^{n,\left( 1\right)}  \right),
\end{align*}
where $\gamma _{n}^{n,\left(2 \right)}=\frac{P_{C}}{\sigma_{n}^{2}}|\tilde{g}_{m,n}|^{2}$, $ \ $ $\tilde{g}_{m,n}=\frac{g_{m,n}}{\sqrt{1+d_{m,n}^{\alpha}}}$, $ \ $ \linebreak $\gamma_{m}^{n,\left(1\right)}= \frac{|h_{m}|^{2}a_{n}^{2}}{|h_{m}|^{2}a_{m}^{2}+\frac{\sigma_{m}^{2}}{P_{BS}}}$ and  $\gamma _{n}^{n,\left(1\right)}= \frac{|h_{n}|^{2}a_{n}^{2}}{|h_{n}|^{2}a_{m}^{2}+\frac{\sigma_{n}^{2}}{P_{BS}}}$.
\subsection*{\textbf{Case 2:} Maintaining Secrecy at both Users}

In this subsection, we study the secure transmission of the messages of both users.  We consider two strategies in which  $U_{m}$ acts  as a relay or acts as a friendly jammer and a relay, simultaneously  (FJR). $U_{n}$ and the eavesdroppers are using the  MRC receivers to decode $S_{n}$.

$U_{m}$ is a \textbf{\textit{relay}}:
In this phase, $U_{m}$ acts as a relay and allocates all of its transmitting power ($P_{C}$) to send the  message of  $U_{n}$. So the transmitted signal of  $U_{m}$  is as \eqref{eq: Direct phase} and the received signals at the eavesdroppers and $U_{n}$ are as:
\begin{align*}
Y_{i}^{\left(2\right)}=\sqrt{P_{C}}\tilde{g}_{m,i}S_{n} +N_{i}^{(2)},
\end{align*}
where $i \in \left\lbrace n,e  \right\rbrace$ and $N_{i}^{(2)}$ is a zero-mean AWGN with variance $\sigma_{i}^{2}$,  $\tilde{g}_{m,e}=\frac{g_{m,e}}{\sqrt{1+d_{m,e}^{\alpha}}}$, such that $g_{m,e}$ and $d_{m,e}$ denote the channel gain and distance between $U_{m}$ and the eavesdroppers, respectively. Now we write the maximum SINR of  an eavesdropper and the SINR of $U_{n}$ to decode $S_{n}$ as shown in the following:
\begin{align*}
\gamma _{i}^{n}=\gamma _{i}^{n,\left(1 \right)}+  \min    \left(  \gamma _{i}^{n,\left(2\right)} \, , \, \gamma _{m}^{n, \left(1 \right)} \right),
\end{align*}
where  $i \in \left\lbrace n,e  \right\rbrace$,  $\gamma _{e}^{n,\left(1\right)}=\frac{|h_{e}|^{2}a_{n}^{2}}{|h_{e}|^{2}a_{m}^{2}+\frac{\sigma_{e}^{2}}{P_{BS}}}$,  $\gamma _{e}^{n,\left(2 \right)}=\frac{P_{C}}{\sigma{e}^{2}} |\tilde{g}_{m,e}|^{2}$.

$U_{m}$ is a \textbf{\textit{FJR}}: In this phase, $U_{m}$  allocates a proportion  $\left( \beta \right)$ of its transmitting power for sending  the message of  $U_{n}$ and the rest of the transmitting power is allocated for sending the noise-like signal. So the transmitted power by $U_{m}$ is as:
\begin{align*}
X_{m}^{\left(  2 \right)}=\left(\sqrt{ \beta} S_{n}+\left( 1-\sqrt{ \beta} \right) X_{J}\right)\sqrt{P_{C}},
\end{align*}
where $X_{J}$ denotes the noise-like signal with power  1. Therefore, the received signals at  the eavesdroppers and $U_{n}$ is obtained as:
\begin{equation*}
\begin{aligned}
Y_{i}^{\left(2 \right)}=\left(\sqrt{ \beta} S_{n}+\left( 1-\sqrt{ \beta} \right) X_{J}\right)\sqrt{P_{C}}\tilde{g}_{m,i}+N_{i}^{(2J)},
\end{aligned}
\end{equation*}
where $i \in \left\lbrace n,e  \right\rbrace$ and $N_{i}^{(2J)}$ is a zero-mean AWGN with variance $\sigma_{i}^{2}$. Now we write the maximum SINR of  an eavesdropper and the SINR of $U_{n}$ as obtained:
\begin{equation*}
\begin{aligned}
\gamma _{i}^{n,\left(2J \right)}=\gamma _{i}^{n,\left(1 \right)}+\min \left( \gamma_{i}^{n,\left(2J\right)} \, , \,\gamma _{m}^{n,\left(1\right)} \right),
\vspace{-1em}
\end{aligned}
\end{equation*}
where $\gamma_{i}^{n,\left(2J\right)}=  \frac{\beta  |\tilde{g}_{m,i}|^{2}P_{C}}{\sigma_{i}^{2}+\left(1- \beta \right)|\tilde{g}_{m,i}|^{2}P_{C}}$.
\section{Secrecy Performance Analysis} We use SOP as the metric to evaluate the secrecy performance of the system for both cases.  For simplicity, we assume that $\sigma _{m}^{2}=\sigma _{n}^{2}=\sigma_{e}^{2}=1$.

\subsection*{\textbf{Case 1:} Maintaining Secrecy at the Stronger User} In this subsection, only the security of the message of  $U_{m}$ is provided , where the targeted data rate of $U_{m}$  is greater than the targeted data rate of $U_{n}$, i.e. , $R_{m} \geq R_{n}$ . As described, when $U_{m}$ is able  to decode $S_{n}$, we have a cooperative NOMA system, otherwise   system goes to outage, i.e. , SOP=1. The outage event $\left( \overline{E} \right)$ occurs if $U_{m}$ can not decode $S_{n} \left(\overline{E}_{1}  \right)$ or $U_{m}$ can not decode $S_{m} \left(  \overline{E}_{2}  \right)$ securely  or $U_{n}$ can not decode $S_{n} \left(\overline{E}_{3}   \right)$.  So we write the outage event as:
\begin{align*}
\text{SOP}=\text{Pr} \left(\overline{E}\right)=1-\text{Pr}\left(E_{1} \cap E_{2} \cap E_{3} \ \right),
\end{align*}
where $E_{1}=\left\lbrace \frac{1}{2} \log \left( 1+\gamma _{m}^{n,\left(  1 \right)} \right) \geq R_{n} \right\rbrace$, \nolinebreak$E_{2}=
$\nolinebreak $\left\lbrace \frac{1}{2} \log \left( \frac{1+\gamma _{m}^{m,\left( 1 \right)}}  {1+\gamma _{e}^{m,\left( 1 \right)}}  \right) \right.$ $\left. \geq R_{m}\right\rbrace$  and $E_{3}=\left\lbrace \frac{1}{2} \log \left( 1+\gamma _{n} ^{n}\right) \geq R_{n}\right\rbrace$.

\begin{theorem}\label{th: independence}
If  $R_{n} \leq R_{m}$, the SOP of the system is as:\\
Condition 1) When $a_{m}^{2} \leq \frac{1}{2^{2R_{n}}+1}$, then:
\begin{align*}
\text{SOP}=1-\text{Pr} \left(  E_{4}  \right) \text{Pr} \left(  E_{2}  \right).
\end{align*}
Condition 2) When $ \frac{1}{2^{2R_{n}}+1}\leq  a_{m}^{2} \leq \frac{1}{2^{2R_{n}}}$, then:
\begin{align*}
 \text{SOP}\geq 1-\text{Pr} \left(  E_{4}  \right) \text{Pr} \left(  E_{2}  \right),
\end{align*}
where $E_{4}=\left\lbrace \frac{1}{2} \log \left(1+\gamma_{n}^{n,\left( 1\right)}+\gamma _{n}^{n,\left(2\right)} \right) \geq R_{n}\right\rbrace.$
\end{theorem}
\begin{IEEEproof}
The proof is provided  in Appendix A.
\end{IEEEproof}
\begin{remark}
 In NOMA systems, $a_{m}^{2}$ must be less than $\frac{1}{2^{2R_{n}}}$ so that the strong user can perform SIC.
\end{remark}
In the following, first we derive $ \text{Pr} \left(  \overline{E}_{2}  \right)$.
\vspace*{-1em}
 \setcounter{equation}{1}
\begin{equation} \label{eq: int E_{4}}
\begin{aligned}
\hspace{-1em}
 \text{Pr} \left(  \overline{E}_{2} \right) \stackrel{(a)}{=}\int\limits _{0}^{\infty}  f_{\gamma_{e}^{m,\left( 1 \right)}} \left(x  \right)  F_{\gamma _{m}^{m,\left( 1\right)}} \left(2^{2R_{m}}\left( 1+ x    \right)-1 \right)    \mathrm{d}x,
\end{aligned}
 \vspace{-1em}
\end{equation}
where $(a)$ holds due to the independence of $\gamma _{e}^{m, \left( 1\right)}$ and $\gamma _{m}^{m, \left( 1\right)}$.
For calculating  $ f_{\gamma_{e}^{m,\left( 1 \right)}} \left(x  \right)$ and $ F_{\gamma _{m}^{m,\left( 1\right)}} \left( x    \right)$ we follow a similar   approach as \cite{liu2017enhancing}.
\begingroup\makeatletter\def\f@size{10}\check@mathfonts
\begin{equation*}
\begin{aligned}
F_{\gamma _{e}^{m,\left(1 \right)}} \left( x \right) =\exp \left[ -\frac{\mu_{1}e^{-\frac{x}{P_{BS}a_{m}^{2}}} \Gamma \left( \eta \, , \, \mu_{2}x   \right)}{x^{\eta}} \right],
\end{aligned}
\end{equation*}
\endgroup
$f_{\gamma _{e}^{m,\left(1 \right)}}\left( x \right)$ is written at the bottom of this page  in \eqref{PDFe}, where
$\eta = \frac{2}{\alpha},\mu_{1} = \eta \pi \lambda _{e}\left( P_{BS}a_{m}^{2}\right)^{\eta}$ and $\mu_{2}=\frac{r_{p}^{\alpha}}{P_{BS}a_{m}^{2}}.$

The $ F_{\gamma _{m}^{m,\left( 1\right)}} \left( x    \right)$ is obtained as [4, eq(4)] :\vspace*{-1em}
 \setcounter{equation}{3}
\begin{equation}\label{eq: CDFm}
\begin{aligned}
&F_{\gamma _{m}^{m,\left( 1\right)}} \left( x \right) = \varphi _{m} \sum _{p=0}^{n_{l} -m}\binom {n_{l}-m}{p} \frac{\left( -1 \right) ^ {p}}{m+p} \\
& \times
\sum _{\tilde{S}_{m}^{p}} \binom {m+p}{q_{0}+ \cdots +q_{N} } \left( \prod _{k=0}^{N} b_{k}^{q_{k}} \right)
e^{- \sum _{k=0}^{N} q _{k}c_{k}\frac{2^{2R_{m}\left(  1+x \right)-1	}}{P_{BS}a_{m}^{2} }},
\vspace{-2em}
\end{aligned}
\end{equation}
\begin{figure*}[!b]
\hrulefill
\begin{align}
\label{eq: Pr E_{2}}
& \text{Pr}\left(  \overline{E}_{2} \right)=\varphi _{m} \sum _{p=0}^{n_{l} -m}\binom {n_{l}-m}{p} \frac{\left( -1 \right) ^ {p}}{m+p}
\sum _{\tilde{S}_{m}^{p}} \binom {m+p}{q_{0}+ \cdots +q_{K} } \left( \prod _{k=0}^{K} b_{k}^{q_{k}} \right) \times  \notag \\
&  \int\limits _{0}^{\infty}  \mu _{1}e^{-\frac{x}{P_{BS}a_{m}^{2}}} \exp \left[ -\frac{\mu_{1}e^{-\frac{x}{P_{BS}a_{m}^{2}}} \Gamma \left( \eta \, , \, \mu_{2}x   \right)}{x^{\eta}} \right]
 \left( \frac{\mu_{2}^{\eta}e^{-\mu_{2}x}}{x}+\frac{\eta  \Gamma \left( \eta \, , \, \mu_{2}x   \right) }{x^{\eta +1}} + \frac{\Gamma \left( \eta \, , \, \mu_{2}x   \right)}{P_{BS}a_{m}^{2} x^{\eta}}\right) e^{- \sum _{k=0}^{K} q _{k}c_{k}\frac{2^{2R_{m}\left(  1+x \right)-1	}}{P_{BS}a_{m}^{2} } } \mathrm{d}x.
\end{align}
\vspace{-1em}
 \begingroup\makeatletter\def\f@size{8.5}\check@mathfonts
 \setcounter{equation}{7}
\begin{align} \label{eq: Pr E_{4}}
&\text{Pr}\left(\overline{E}_{4}\right) =\text{Pr} \left\lbrace  \frac{1}{2} \log \left( 1+ \gamma _{n}^{n,\left( 1 \right)} +\gamma _{n}^{n,\left( 2\right)} \right) < R_{n}  \right\rbrace =\int\limits _{0}^{\infty}   f_{\gamma _{n}^{n,\left(2  \right)}} \left( x \right)    F_{\gamma _{n}^{n,\left(   1\right)}}\left( 2^{2R_{n}}-1-x\right) \, \mathrm{d}x =\notag \\
&-\frac{2 \lambda_{m,n}}{\pi P_{C}} \sum\limits _{k=1}^{N} B_{k} C_{k}\int\limits _{0}^{\infty} e^{-C_{k} \frac{\lambda _{m,n}x}{P_{C}}} \left(
U(2^{2R_{n}}-1-x-\theta)+ U(-2^{2R_{n}}+1+x+\theta) \varphi _{n} \sum _{p=0}^{n_{l} -n}\binom {n_{l}-n}{p} \frac{\left( -1 \right) ^ {p}}{n+p}\right.  \notag \\
&\times \left. \sum _{\tilde{S}_{n}^{p}} \binom {n+p}{q_{0}+ \cdots +q_{N} } \left( \prod _{k=0}^{N} b_{k}^{q_{k}} \right)
 e^{- \sum _{k=0}^{N} q _{k}c_{k}\frac{\left(2^{2R_{n}}-1-x \right) }{\left(a_{n}^{2} -a_{m}^{2}\left( 2^{2R_{n}}-1-x \right)   \right)P_{BS}}} \right)\mathrm{d}x.
\end{align}
\endgroup
\end{figure*}
where  $N$ is  a parameter defined to guarantee the complexity-accuracy trade off,  $\varphi _{m}= \frac{n_{l}!}{\left( n_{l} - m \right)! \left(  m-1\right)!}$ and $b_{k} = -\frac{\omega _ {N}}{2} \sqrt{1-\phi _{k}^{2}} \left( \phi_{k} + 1 \right), \  b_{0}= -\sum_{k=1}^{N} b_{k}$, \linebreak$ c_{k}=1+ \left[ \frac{r_{l}}{2} \left( \phi_{k} + 1 \right) \right]^{\alpha} , \,   c_{0}=0, \,  \omega _{N}=\frac{\pi}{N} , \  \phi _{k} = \cos{ \left( \frac{2k-1}{2N} \pi \right)},$  $\tilde{S}_{m}^{p}= \left\lbrace \left( q_{0}+ \cdots +q_{N} \right) | \sum _ {i=0}^{N} q _ {i} = m+ p \right\rbrace$ and $\binom{m+p}{q_{0}+ \cdots +q_{N}}=\frac{\left( m+p\right)!}{ q_{0}! \cdots q_{N}! }.$ Now by substituting \eqref{PDFe} and \eqref{eq: CDFm} into \eqref{eq: int E_{4}}, $\text{Pr} \left( \overline{E}_{2}  \right)$ is derived as illustrated at the bottom of this page in \eqref{eq: Pr E_{2}}.

 The last step for finding the SOP is deriving $\text{Pr} \left(  \overline{E}_{4}\right).$ The  $F_{\gamma _{n}^{n,\left(1 \right)}}$  is obtained as [4, eq(5)].
 \vspace{-1em}
\begin{equation*}
\begin{aligned}
&F_{\gamma _{n}^{n,\left( 1\right)}} \left( x \right) = \varphi _{n} \sum _{p=0}^{n_{l} -n}\binom {n_{l}-n}{p} \frac{\left( -1 \right) ^ {p}}{n+p}  \sum _{\tilde{S}_{n}^{p}} \binom {n+p}{q_{0}+ \cdots +q_{N} } \\ & \times \left( \prod _{k=0}^{N} b_{k}^{q_{k}} \right)
e^{- \sum _{k=0}^{N} q _{k}c_{k}\frac{2^{2R_{n}\left(  1+x \right)-1	}}{P_{BS}\left(a_{n}^{2}-a_{m}^{2} \left(1+x   \right)-1   \right) }} U \left( -x+\theta  \right) +\\
&U \left( x-\theta  \right),
\end{aligned}
\end{equation*}
 where $\theta=\frac{a_{n}^{2}}{a_{m}^{2}},$ $\varphi _{n}= \frac{n_{l}!}{\left( n_{l} - n \right)! \left(  n-1\right)!}$,  $\tilde{S}_{n}^{p}= \left\lbrace \left( q_{0}+ \cdots +q_{N} \right) | \sum _ {i=0}^{N} q _ {i} = n+ p \right\rbrace $  and $\binom{n+p}{q_{0}+ \cdots +q_{N}}=\frac{\left( n+p\right)!}{ q_{0}! \cdots q_{N}! }.$

 According to \cite{moltchanov2012distance}, $d_{m,n}$ is a random variable with the probability density function as:
 \setcounter{equation}{5}
\begin{equation}
\begin{aligned}
\label{eq: pdf users}
&f_{d_{m,n}}\left( r \right)= \frac{2r}{r_{l}^{2}} \left( \frac{2}{\pi} \cos ^{-1} \left(\frac{r}{2r_{l}} \right) -\frac{r}{\pi r_{l} } \sqrt{1-\frac{r^{2}}{4r_{l}^{2}}} \right),
\end{aligned}
\end{equation}
where $0\leq r \leq 2r_{l}$. By using Gaussian-Chebyshev quadrature method, $F_{\gamma _{n}^{n,\left( 2 \right)}} \left(y \right)$  and $f_{\gamma _{n}^{n,\left( 2 \right)}} \left(y \right)$ are derived  as the method  in \cite{ding2014performance}. The details of the proof are provided  in Appendix B.
\begin{equation}
\hspace{-1em}
\begin{aligned}
\label{eq: CDFn2}
F_{\gamma _{n}^{n,\left( 2 \right)}} \left( y \right) \approx \frac{2}{\pi} \sum\limits _{k=0}^{N} B_{k} \, e^{-C_{k} \frac{\lambda _{m,n}y}{P_{C}}},
\end{aligned}
\vspace{-1em}
\end{equation}
\begin{equation*}
\vspace*{-1em}
\begin{aligned}
 f_{\gamma _{n}^{n,\left( 2 \right)}} \left( y \right) \approx \frac{-2 \lambda_{m,n}}{\pi P_{C}} \sum\limits _{k=1}^{N} B_{k}C_{k} \, e^{-C_{k} \frac{\lambda _{m,n}y}{P_{C}}},
\end{aligned}
\end{equation*}
where  $B_{k}= -\omega _{N} \sqrt{1-\theta _{k}^{2}}  \left( 1+\theta _{k}  \right)  \left( 2 \cos ^{-1} \left(\frac{\left( 1+\theta _{k}  \right)}{2} \right)- \right.$
$ \left. \left( 1+\theta _{k}  \right) \sqrt{1-\frac{\left( 1+\theta _{k}  \right)^{2}}{4}} \right) ,$ $B_{0}=-\sum\limits_{k=1}^{N}B_{k},$ $\omega_{N}=\frac{\pi}{N},$ \linebreak $C_{0}=0,$ $C_{k}=  1+\left( r_{l}+r_{l} \theta _{k}  \right)^{\alpha}$ and $\theta _{k}= \cos \left( \frac{2k-1}{2N} \pi \right)$. Due to the independence of the $|h_{n}|^{2}$ and $|\tilde{g}_{m,n}|^{2}$,   $ \text{Pr}\left(\overline{E}_{4} \right)$  is derived as  the bottom of  this page in \eqref{eq: Pr E_{4}}.
\vspace*{-1em}
\subsection*{\textbf{Case 2:} Maintaining Secrecy at both Users}
\label{sec: both}
In this subsection, the messages of two users are sent securely, where the targeted data rates of $U_{m}$ and $U_{n}$ are $R_{m}$ and $R_{n}$, respectively.  The only effect of cooperation on $U_{m}$ is to divide the transmission duration by 2, thus in this subsection, we investigate the SOP of $U_{n}$ only and for avoiding untractable calculations, we derive an upper bound on the SOP of $U_{n}$ at high SNR regimes as \cite{chen2018physical}  and \cite{bassem2018passive}, where \linebreak $P_{BS}\rightarrow \infty$. We  remark that at high SNR regimes, on the condition $a_{m}^{2} \leq \frac{1}{2^{2_{R_{n}}}}$ , $U_{m}$ is able to carry out the SIC with probability one. So the SOP of the system is obtained as:
\begin{equation*}
\begin{aligned}
\text{SOP}=1-\left( 1-\text{SOP}_{m}\right)\left( 1-\text{SOP}_{n}\right),
\end{aligned}
\end{equation*}
\setcounter{equation}{8}
\begin{figure*}[!b]
\hrulefill
 \begin{equation}
 \begin{aligned}
 \label{eq:PDFeJ}
 &f_{\gamma _{e}^{n,\left( 2J \right)}}\left( x \right)=\chi_{2}\beta P_{C}e^{-\left(\frac{x}{ \beta P_{C}-\left( 1-\beta \right) P_{C}x }+\chi_{2} e^{-\frac{x}{\beta P_{C} -\left(1-\beta  \right)P_{C}x}} \left(  \frac{\beta P_{C} -\left(1-\beta  \right)P_{C}\,x}{x}  \right) ^{\eta} \,\right)}
  \left(\frac{\left( \beta P_{C} -\left(1-\beta  \right)P_{C}x  \right)^{\eta -2}}{x^{\eta}}\right. \\
  &\left. +\eta \frac{\left( \beta P_{C} -\left(1-\beta  \right)P_{C}x  \right)^{\eta -1}}{x^{\eta +1}}\right) U\left(  -x+\frac{\beta}{1-\beta}\right).
 \end{aligned}
 \end{equation}
\end{figure*}
where $\text{SOP}_{m}$ equals to $\text{Pr} \left( \overline{E}_{2}\right)$ which is derived in \eqref{eq: Pr E_{2}}. So we write $\text{SOP}_{n}$ at high SNR regimes  as:
 \begingroup\makeatletter\def\f@size{8.5}\check@mathfonts
\begin{align*}
\text{SOP}_{n}=\text{Pr} \left\lbrace  \frac{1+ \theta +\min \left(\gamma _{n}^{n, \left(  T \right)} \, , \, \theta  \right)}{1+\theta + \min \left(\gamma _{e}^{n,\left(T \right)} \, ,\, \theta \right) }  <C_{n}^{g} \right\rbrace,
\end{align*}
\endgroup
where $C_{n}^{g}=2^{2R_{n}}$ and $T$ is used to differentiate between two cases, the relay (shown by $T=2$) and the  FJR (shown by $T=2J$).
\subsubsection{$U_{m}$ is a relay }
In this subsection, $U_{m}$ acts as a relay and allocates all of the $P_{C}$ to send $S_{n}$. SOP of $U_{n}$ is as the following lemma. (Proof is provided in  Appendix C.)
\begin{lemma} \label{lemma: both users}
Let  $\zeta=\left(  C_{n}^{g}-1\right) \left(1+\theta \right) $,  if  $\theta \leq \zeta$, then \linebreak  $\text{SOP}_{n}=1$, otherwise we have:
 \begingroup\makeatletter\def\f@size{8.5}\check@mathfonts
\begin{equation*}
\begin{aligned}
\text{SOP}_{n}=1-F_{ \gamma _{e}^{n,\left(  2\right)} } \left(\frac {\theta -\zeta}{C_{n}^{g}}  \right)+\int\limits_{0}^{\frac{\theta -\zeta}{C_{n}^{g}}}  f_{ \gamma _{e}^{n,\left(  2\right)}} \left(y \right) F_{ \gamma _{n}^{n,\left(  2\right)}} \left( \zeta + C_{n}^{g} y  \right) \mathrm{d}y.
\end{aligned}
\end{equation*}
\endgroup
\end{lemma}
Now we calculate the $F_{\gamma _{e}^{n,\left(2 \right)}}\left(x\right)$ and   $f_{\gamma _{e}^{n,\left(2 \right)}}\left(x\right)$ in order to derive the $\text{SOP}_{n}$, by following a similar way as \cite{liu2017enhancing}.
\begin{lemma} \label{lemma : gamma e}
Let  $\chi_{1}= \pi \eta \lambda _{e} P_{C}^{ \eta}$. The  $F_{\gamma _{e}^{n,\left(2 \right)}}\left(x\right)$ and  $f_{\gamma _{e}^{n,\left(2 \right)}}\left(x\right)$ are as following: (Proof is provided in  Appendix D.)
 \begingroup\makeatletter\def\f@size{10}\check@mathfonts
\begin{equation*}
\begin{aligned}
F_{\gamma _{e}^{n,\left(2 \right)}}\left(x\right)=e^ { -\chi_{1} \frac{e^{-\frac{x}{P_{C}}}\Gamma \left(  \eta   \right)}{x^{\eta}}},
\end{aligned}
\vspace{-1.5em}
\end{equation*}
\begin{align*}
f_{\gamma _{e}^{n,\left(2\right)}} \left(  x \right)= \chi _{1} \Gamma \left( \eta \right)e^ { -\left(\frac{\chi_{1}e^{-\frac{x}{P_{C}}}\Gamma \left(  \eta   \right)}{x^{\eta}}+\frac{x}{P_{C}}\right)} \left( \frac{\eta}{x^{\eta +1}} + \frac{1}{P_{C}x^{\eta}}\right).
\end{align*}
\endgroup
\end{lemma}
\vspace*{-1em}
\subsubsection{$U_{m}$ is a FJR }
In this subsection, we investigate the  $\text{SOP}_{n}$ at high SNR regimes, while $U_{m}$ is acting as a friendly jammer and a relay, simultaneously. So  we derive the  $\text{SOP}_{n}$ by using the following lemma. (Proof is provided in  Appendix C.)
\begin{lemma} \label{lemma: both users Jam}
If $\theta \leq \zeta$, then  $\text{SOP}_{n}=1,$ otherwise we have:
 \begingroup\makeatletter\def\f@size{10}\check@mathfonts
\begin{equation*}
\begin{aligned}
&\text{SOP}_{n}=1-F_{ \gamma _{e}^{n,\left(2J \right)}} \left(\frac {\theta -\zeta}{C_{n}^{g}}  \right)\\&+\int\limits_{0}^{\frac{\theta -\zeta}{C_{n}^{g}}}  f_{ \gamma _{e}^{n,\left(2J \right)}} \left(y \right) F_{ \gamma _{n}^{n,\left(2J \right)}} \left( \zeta + C_{n}^{g} y  \right) \mathrm{d}y.
\end{aligned}
\vspace{-0.5em}
\end{equation*}
\endgroup
\end{lemma}
Now we find the terms of the $\text{SOP}_{n}$ as obtained in the following.
\begin{lemma} \label{lemma: gamma eJ}
 Let $\chi_{2}=\pi \eta \Gamma \left(\eta \right) \lambda _{e}$, the  $F_{\gamma _{e}^{n,\left( 2J \right)}} \left(x \right)$ and equal to:  (Proof is provided in Appendix E.)
 \begingroup\makeatletter\def\f@size{9}\check@mathfonts
 \begin{equation*}
 \begin{aligned}
& F_{ \gamma _{e}^{n,\left( 2J \right)}}\left( x \right)=U \left(  x-\frac{\beta}{1-\beta} \right)+U \left( -x+\frac{\beta}{1-\beta} \right)\times   \\
& \exp \left[  -\chi_{2} e^{-\frac{x}{\beta P_{C} -\left(1-\beta  \right)P_{C}\,x}} \left(  \frac{\beta P_{C} -\left(1-\beta  \right)P_{C}\,x}{x}  \right) ^{\eta} \, \right] ,
 \end{aligned}
 \end{equation*}
 \endgroup
and  $ f_{ \gamma _{e}^{n,\left( 2J \right)}}\left( x \right)$ is written at the bottom of this page in  \eqref{eq:PDFeJ}.
\end{lemma}

By following a similar approach as \eqref{eq: CDFn2} and using Gaussian-Chebyshev quadrature method, the  $F_{\gamma _{n}^{n,\left(n,2J\right) }} \left( x\right)$  is derived as: \begingroup\makeatletter\def\f@size{8.5}\check@mathfonts
\begin{equation*}
\begin{aligned}
&F_{\gamma _{n}^{n,\left(n,2J\right) }} \left( x\right)= \text{U} \left(x- \frac{\beta}{1-\beta} \right)\\
&+ \text{U} \left(-x+ \frac{\beta}{1-\beta} \right)  \frac{2}{\pi} \sum\limits _{k=0}^{N} B_{k} \, e^{-C_{k}\frac{\lambda _{m,n}x}{P_{C} \left( \beta - \left(1-\beta \right)x \right)}}.
\end{aligned}
\end{equation*}
\endgroup
\section{Numerical Results}
\begin{table}
\centering
\caption{Parameters Of  The  Simulations}
 \label{table: Parameters}
\begin{tabular}{|l|l|}
\hline
Number of iterations in Monte Carlo simulations  & $10^{5}$  \\
\hline
The radius of the eavesdropper zone & $r_{e}=100m$  \\
\hline
The radius of the  user zone & $r_{l}=10m$  \\
\hline
The radius of  the eavesdropper-free zone & $r_{p}=5m$  \\
\hline
Path-loss exponent & $\alpha=4$  \\
\hline
Order of the users & $n_{l}=2, n=1, m=2$  \\
\hline
Users' targeted data rates & $R_{n}=R_{m}=0.1$  \\
\hline
The power allocation coefficients & $a_{n}^{2}=0.6 , a_{m}^{2}=0.4$  \\
\hline
The density of the eavesdroppers & $\lambda_{e}=10^{-3}$  \\
\hline
The complexity-vs-accuracy coefficient & $N=20$  \\
\hline
\end{tabular}
\end{table}
\begin{figure}[tb]
\centering
\includegraphics[width=7.5cm,height=6cm]{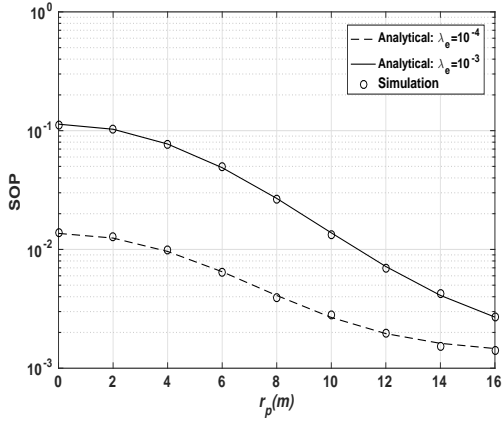}
\caption{SOP of the system versus $r_{p}$ for different $\lambda_{e}$ values with: $ P_{C}=20\text{dB}$ and $P_{BS}=60\text{dB}.$}
\end{figure}
In this section, we present numerical and simulation results, where the parameters of the simulations are shown in the Table \ref{table: Parameters}. 

\subsection*{\textbf{Case 1:} Maintaining Secrecy at the Strong User}

Fig. 2 illustrates that SOP of the system decreases by increasing the radius of the eavesdropper-free zone $\left(r_{p}\right)$, thanks to increasing the distance of the eavesdroppers. Also as expected, by increasing $\lambda_{e}$ SOP of the system increases and  the simulation results confirm our analytical results.
\begin{figure}[tb]
\centering
\hspace{0.5em}
\includegraphics[width=7.19cm,height=6cm]{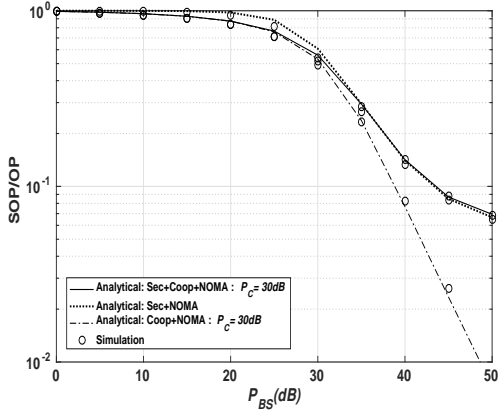}
\caption{Comparison of the SOP of  the cooperative NOMA systems with the eavesdroppers (Sec+Coop+NOMA), cooperative NOMA systems without the eavesdroppers (Coop+NOMA) and NOMA systems with  the eavesdroppers (Sec+NOMA).  }
\vspace{-2em}
\end{figure}

Fig. 3 shows the comparison of  NOMA systems and cooperative NOMA systems with and without the  eavesdroppers. We see that at high SNR regimes, the SOP of the system doesn't vary by increasing  $P_{BS}$. Since at high SNR regimes, the outage probability of the weak user is almost zero and the SOP of the system equals to the SOP of the strong user which is almost independent of $P_{BS}$ and only depends on $|h_{m}|^{2}$ and $|h_{e}|^{2}$. This is due to the increase occurred in $\rho_{e}$ by increasing $P_{BS}$.  So it depends on the users' targeted data rates and power allocation coefficients in order to determine the usefulness of the  cooperation. AS observed in Fig. 3, for lower $P_{BS}$, Coop+NOMA+Sec outperforms NOMA+Sec and thus cooperation is beneficial, while for high $P_{BS}$ using the cooperation degrades the performance of the system. This is due to the half duplex property of the relay, in which half of the time resource is allocated to relaying in cooperative NOMA and thus the rate of the strong user is divided by 2.
\subsection*{\textbf{Case 2:} Maintaining Secrecy at both users }
When we have secrecy at both users, the SOP of the  strong user is the same as the case of the maintaining secrecy at the strong user and  therefore we only investigate the SOP of the weak user, while the strong user is a relay or a FJR, at high SNR regimes $\left(P_{BS}\rightarrow \infty \right)$. As Fig. 4 indicates, for $P_{C}\rightarrow \infty$, the SOP of  $U_{n}$ goes to one. Since by increasing $P_{C}$, the received power at the $U_{n}$ and eavesdroppers increases and therefore the eavesdroppers would be able to decode $S_{n}$ for $P_{C}\rightarrow \infty $ with probability  one. For the lower  $P_{C}$, the increment of the received power at  $U_{n}$ is greater than   the increment of the received power at the eavesdroppers, thus the SOP of the $U_{n}$ decreases by increasing $P_{C}$. When $\beta$ is very close to zero, $U_{n}$ would not be able to decode the $S_{n}$. Moreover, we see that for $\beta=0.7$,  the FJR strategy has a better secrecy performance than the relaying strategy  for the high value of $P_{C}$. Besides, at the low value of $P_{C}$, it is better to choose the relaying instead of  FJR strategy, for  the given users' targeted data rates and power allocation coefficients.
\begin{figure}[tb]
\centering
\includegraphics[width=7cm,height=6cm]{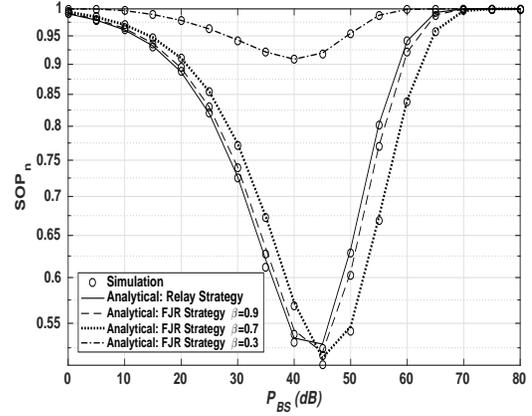}
\caption{SOP of the $U_{n}$(weak user) for different $\beta$ values.}
\vspace{-1.5em}
\end{figure}
\section{Conclusion}
We studied the secrecy performance of a cooperative  NOMA system with many legitimate users  in existence of a random number of external passive eavesdroppers in two cases: either security of the strong user or both users were provided, while the strong user was a relay or a friendly jammer. In case 1, we derived a lower bound on the SOP which is tight in some regions of the power allocation coefficients and users' targeted data rates.  In case 2, we derived an upper bound on the SOP of the system at high SNR regimes. Our results showed that the amount of  power must be allocated to send jamming noise has an optimal value that might be derived in a future work.

\onecolumn
\appendices
 \section{Proof Of Theorem 1}
  First, we provide a lemma that we use it for deriving the SOP the system for the case of maintaining the secrecy at the strong user.
  \begin{lemma}\label{lemma: independence}
If $R_{n}\leq R_{m}$ and $a_{m}^{2} \leq \frac{1}{2^{2R_{n}}+1}$, then we have  $E_{2} \subseteq E_{1}.$
\end{lemma}
\begin{IEEEproof}
We use proof by contradiction. Our contradiction assumption is :
\begin{align*}
&\frac{1}{2} \log \left(\frac{1+|h_{m}|^{2}a_{m}^{2}P_{BS}}{1+|h_{e}|^{2}a_{m}^{2}P_{BS}} \right) \geq R_{m},
\end{align*}
which implies that:
\begin{align}\label{eq: lower band hm}
& |h_{m}|^{2}a_{m}^{2}P_{BS} \geq 2^{2R_{m}} |h_{e}|^{2}a_{m}^{2}P_{BS} +\left(2^{2R_{m}}-1\right).
\end{align}
Also, on the condition $\frac{a_{n}^{2}}{a_{m}^{2}} \geq  2^{2R_{n}}-1:$
\begin{align*}
&\frac{1}{2} \log \left( 1+\frac{|h_{m}|^{2}a_{n}^{2}}{|h_{m}|^{2}a_{m}^{2}+\frac{1}{P_{BS}}}\right) < R_{n},
\end{align*}
which implies that:
\begin{align}\label{eq: upper band hm}
& |h_{m}|^{2} <\frac{ 2^{2R_{n}}-1}{\left( a_{n}^{2}-a_{m}^{2}\left( 2^{2R_{n}}-1 \right)  \right) P_{BS}}.
\end{align}
By substituting \eqref{eq: upper band hm} into \eqref{eq: lower band hm} we have:
\begin{align} \label{eq: contrast}
\frac{\left( 2^{2R_{n}}-1\right)a_{m}^{2}}{\left( a_{n}^{2}-a_{m}^{2}\left( 2^{2R_{n}}-1 \right)  \right) } > 2^{2R_{m}} |h_{e}|^{2}a_{m}^{2}P_{BS} +\left(2^{2R_{m}}-1\right).
\end{align}
By using the assumptions $R_{n}\leq R_{m}$ and $a_{m}^{2} \leq \frac{1}{2^{2R_{n}}+1},$ we know $2^{2R_{m}}-1 >\frac{\left( 2^{2R_{n}}-1\right)a_{m}^{2}}{\left( a_{n}^{2}-a_{m}^{2}\left( 2^{2R_{n}}-1 \right)  \right) }$, which is in contrast to \eqref{eq: contrast}.
So we are sure that if $E_{2}$ occurs then $E_{1}$ will occur. Therefore, $E_{2} \subseteq E_{1}.$
\end{IEEEproof}
 \label{appendix: theorem}
SOP of the system is as:
\begin{align} \label{eq: Final SOP definition}
\text{SOP}=\text{Pr} \left(\overline{E}\right)=1-\text{Pr}\left(E_{1} \cap E_{2} \cap E_{3} \ \right).
\end{align}
$a_{m}^{2}$ must be less than $\frac{1}{2^{2R_{n}}}$ so that $U_{m}$ can carry out SIC and decode $S_{n}.$ Now we find  $\text{Pr} \left(E_{1}\cap E_{2} \cap E_{3} \right)$ as it is shown.
\begingroup\makeatletter\def\f@size{8.5}\check@mathfonts
\begin{align*}
 &\text{Pr} \left(E_{1}\cap E_{2} \cap E_{3} \right)=
\text{Pr} \left\lbrace \frac{1}{2} \log \left( 1+\gamma _{n}^{n,\left(1 \right)}+\min \left(\gamma _{n}^{n,\left( 2 \right)} \, , \, \gamma _{m}^{n,\left(1\right)} \right) \right) \geq R_{n} 
  \cap  \frac{1}{2} \log \left( 1+\gamma _{m}^{n,\left( 1\right)}   \right) \geq R_{n} \cap E_{2} \right\rbrace =\\
 &\text{Pr} \left\lbrace \frac{1}{2} \log \left(1+\gamma_{n}^{n,\left( 1\right)}+\gamma _{n}^{n,\left(2\right)} \right) \geq R_{n}
 \cap  \frac{1}{2} \log \left(1+\gamma_{n}^{n,\left( 1\right)}+\gamma _{m}^{n,\left(1\right)} \right) \geq R_{n} \cap  \frac{1}{2} \log \left( 1+\gamma _{m}^{n,\left( 1\right)}   \right) \geq R_{n} \cap E_{2} \right\rbrace = \\
&\text{Pr}  \left\lbrace \frac{1}{2} \log \left(1+\gamma_{n}^{n,\left( 1\right)}+\gamma _{n}^{n,\left(2\right)} \right) \geq R_{n} \cap   \frac{1}{2} \log \left( 1+\gamma _{m}^{n,\left( 1\right)}   \right) \geq R_{n} \cap E_{2} \right\rbrace.
\end{align*}
\endgroup
By using lemma \ref{lemma: independence} and the fact that the  channel coefficients are independent  (assume $a_{m}^{2} \leq \frac{1}{2^{2R_{n}}+1}$) we have:
\begingroup\makeatletter\def\f@size{8.5}\check@mathfonts
\begin{align} \label{eq: E_{4}}
 &\text{Pr} \left(E_{1}\cap E_{2} \cap E_{3} \right)= 
 \text{Pr}    \left\lbrace\frac{1}{2} \log \left(1+\gamma_{n}^{n,\left( 1\right)}+\gamma _{n}^{n,\left(2\right)} \right) \geq R_{n}\right\rbrace \text{Pr}  \left\lbrace \frac{1}{2} \log \left( 1+\gamma _{m}^{n,\left( 1\right)}   \right) \geq R_{n} \cap E_{2} \right\rbrace= 
 \text{Pr} \left( E_{4} \right) \text{Pr} \left(  E_{2} \right).
\end{align}
\endgroup
By substituting  \eqref{eq: E_{4}} into \eqref{eq: Final SOP definition},  SOP of the system is as:
\begin{align*}
\text{SOP}= 1-\text{Pr} \left( E_{4}  \right) \text{Pr} \left(  E_{2}  \right).
\end{align*}
If the conditions $\frac{1}{2^{2R_{n}}} \leq a_{m}^{2} \leq \frac{1}{2^{2R_{n}}+1},$ $ \text{Pr}  \left\lbrace \frac{1}{2} \log \left( 1+\gamma _{m}^{n,\left( 1\right)}   \right) \geq R_{n} \cap E_{2} \right\rbrace \leq \text{Pr} \left(E_{2} \right) $ hold, then the following lower bound holds  on the SOP of the system:
\begin{align*}
\text{SOP}\geq 1-\text{Pr} \left( E_{4}  \right) \text{Pr} \left(  E_{2}  \right).
\end{align*}
\section{ CDF and PDF Of    $\gamma_{n}^{n,\left( 2 \right)}$}
\label{appendix: gamma2}
For deriving the  $F_{\gamma_{n}^{n,\left( 2 \right)}}\left(y \right)$ and $f_{\gamma_{n}^{n,\left( 2 \right)}}\left(y \right)$ , we follow a similar way as \cite{ding2014performance}. According to the exponential distribution of the $|g_{m,n}|^{2}$,  \eqref{eq: pdf users} and also due to the independence of the $d_{m,n}$ and $|g_{m,n}|^{2}$, the  $F_{\gamma_{n}^{n,\left( 2 \right)}}\left(y \right)$ is computed as:
\begingroup\makeatletter\def\f@size{9}\check@mathfonts
\begin{equation*}
\begin{aligned}
F_{\gamma _{c}} \left( y \right) &=
\text{Pr} \left\lbrace  P_{C} \frac{|g_{m,n}|^{2}}{1+d_{m,n}^{\alpha}} \leq y  \right\rbrace =
\text{Pr} \left\lbrace   |g_{m,n}|^{2} \leq \frac{\left(1+d_{m,n}^{\alpha}\right) y}{P_{C}}  \right\rbrace 
=\int\limits _{0} ^{2r_{l}} f_{d_{m,n}} \left(  r \right) F_{ |g_{m,n}|^{2}} \left(   \frac{\left(1+r^{\alpha}\right) y}{P_{C}}  \right) \mathrm{d}r\\
&=\int\limits_{0}^{2r_{l}} \frac{2r}{r_{l}^{2}} \left( \frac{2}{\pi} \cos ^{-1} \left(\frac{r}{2r_{l}} \right) -\frac{r}{\pi r_{l} } \sqrt{1-\frac{r^{2}}{4r_{l}^{2}}} \right) \left(1- e^{-\lambda_{m,n}\left(  1+r^{\alpha}  \right)\frac{y}{P_{C}}}   \right) \mathrm{d}r ,
\end{aligned}
\end{equation*}
\endgroup
now by using Gaussian-Chebyshev quadrature method \cite{ding2014performance}, we have:
\begin{align*}
F_{\gamma _{n}^{n,\left(2 \right)}} \left( y \right) \approx \frac{2}{\pi} \sum\limits _{k=0}^{N} B_{k} \, e^{-C_{k} \frac{\lambda _{m,n}y}{P_{C}}}.
\end{align*}
Taking the derivative of  $F_{\gamma _{n}^{n,\left(2 \right)}} \left( y \right) $, we find  $f_{\gamma _{n}^{n,\left(2 \right)}} \left( y \right) $ as:
\begin{align*}
f_{\gamma _{n}^{n,\left( 2 \right)}} \left( y \right) \approx \frac{-2 \lambda_{m,n}}{\pi P_{C}} \sum\limits _{k=1}^{N} B_{k}C_{k} \, e^{-C_{k} \frac{\lambda _{m,n}y}{P_{C}}}.
\end{align*}
\section{Proof  Of  Lemmas \ref{lemma: both users} And \ref{lemma: both users Jam}}
 \label{appendix: lemma}
 We prove the lemmas \ref{lemma: both users} and \ref{lemma: both users Jam} at the same time by using index " $ T $ " to differentiate between two strategies of relaying ($T=2$)  and FJR ($T=2J$). We follow a similar way as \cite{bassem2018passive}, so we have:
\begin{align}\label{eq: SOP both}
&\text{SOP}_{n}=\text{Pr} \left\lbrace  \frac{1+ \theta +\min \left( \gamma _{n}^{n,\left( T\right)} \, , \, \theta  \right)}{1+\theta + \min \left( \gamma _{e}^{n,\left( T\right)} \, ,\, \theta \right) }  <C_{n}^{g} \right\rbrace 
=1- \text{Pr}\underbrace { \left\lbrace \min \left( \gamma _{n}^{n,\left( T\right)} \, , \, \theta    \right)   \geq \zeta + C_{n}^{g} \min \left( \gamma _{e}^{n,\left( T\right)}\, , \, \theta  \right) \right\rbrace}_{F} ,
\end{align}
when  $F$ is written as a union of four distinct events $F_{1},F_{2},F_{3}$ and $F_{4}$ $\left( F=\bigcup\limits _{i=1}^{4} F_{i}  \right)$ such that $F_{1}=\left\lbrace \gamma _{n}^{n,\left( T\right)} \geq \theta  \cap \ \theta \geq \zeta +C_{n}^{g} \, \gamma _{e}^{n,\left( T\right)}\right\rbrace,$ $F_{2}=\left\lbrace \gamma _{n}^{n,\left( T\right)} \geq \theta \ \cap \right.$ $\left. \ \theta \geq \zeta +C_{n}^{g} \theta \right\rbrace  $,  $F_{3}=\left\lbrace \gamma _{n}^{n,\left( T\right)} < \theta \ \cap \  \gamma _{n}^{n,\left( T\right)} \geq \zeta +C_{n}^{g} \theta \right\rbrace$ and $F_{4}=\left\lbrace \gamma _{n}^{n,\left( T\right)} < \theta \ \cap \  \gamma _{n}^{n,\left( T\right)} \geq \zeta +C_{n}^{g} \, \gamma _{e}^{n,\left( T\right)}\right\rbrace$. So we have:
\begin{align*}
 \text{Pr} \left( F\right) =\text{Pr} \left( F_{1} \right) +  \text{Pr} \left(F_{2}  \right) +   \text{Pr} \left(F_{3} \right) + \text{Pr} \left(F_{4} \right).
\end{align*}
Since $C_{n}^{g} \geq 1$, then  $\text{Pr} \left( F_{2}\right)=\text{Pr} \left( F_{3} \right)=0 $ and thus:
\begin{align}\label{eq: Final Pr F}
 \text{Pr} \left( F\right) =\text{Pr} \left( F_{1} \right) +\text{Pr} \left(F_{4} \right).
\end{align}
In the following, first we derive the $\text{Pr} \left( F_{1} \right)$. Since $\gamma _{n}^{n,\left(T\right)}$  and $\gamma _{e}^{n,\left( T\right)} $ are independent, so we have:
\begin{align}\label{eq: Pr F_{1}}
\text{Pr} \left\lbrace F_{1}  \right\rbrace =
\text{Pr} \left\lbrace   \gamma _{n}^{n,\left( T\right)} \geq \theta    \right\rbrace \text{Pr} \left\lbrace  \theta \geq \zeta +C_{n}^{g} \,  \gamma _{e}^{n,\left( T\right)} \right\rbrace  
=\left(1- F_{ \gamma _{n}^{n,\left( T\right)}} \left(  \theta  \right) \right) F_{ \gamma _{e}^{n,\left( T\right)} } \left(\frac {\theta -\zeta}{C_{n}^{g}}  \right).
\end{align}
For deriving $\text{Pr} \left(F_{4}\right)$, on the condition $  \theta >\zeta$, we have:
 \begingroup\makeatletter\def\f@size{9}\check@mathfonts
\begin{align}\label{eq: Pr F_{4}}
\text{Pr} \left(F_{4}\right)& =
\int\limits_{0}^{\frac{\theta -\zeta}{C_{n}^{g}}}  f_{\gamma _{e}^{n,\left( T\right)}} \left(y \right) \int\limits_{\zeta + C_{n}^{g} y} ^{\theta} f_{ \gamma _{n}^{n,\left( T\right)}} \left( x \right) \, \mathrm{d}x \, \mathrm{d}y= 
\int\limits_{0}^{\frac{\theta -\zeta}{C_{n}^{g}}}  f_{ \gamma _{e}^{n,\left(  T\right)}} \left(y \right) \left( F_{ \gamma _{n}^{2,\left(  T\right)}} \left(    \theta \right)-F_{\gamma _{n}^{n,\left(  T\right)}} \left( \zeta + C_{n}^{g} y  \right)\right) \mathrm{d}y   \notag \\
&= F_{\gamma _{e}^{n,\left(  T\right)}} \left(\frac{\theta -\zeta}{C_{n}^{g}}   \right) F_{\gamma _{n}^{n,\left(  T\right)}} \left(    \theta \right) -
\int\limits_{0}^{\frac{\theta -\zeta}{C_{n}^{g}}}  f_{ \gamma _{e}^{n,\left(  T\right)}} \left(y \right) F_{ \gamma _{n}^{n,\left(  T\right)}} \left( \zeta + C_{n}^{g} y  \right) \mathrm{d}y .
\end{align}
\endgroup
Finally by substituting ( \ref{eq: Final Pr F} - \ref{eq: Pr F_{4}}) into \eqref{eq: SOP both},  under the condition $\theta \leq \zeta,$ $\text{SOP}_{n}=1$. Otherwise, we have:
 \begingroup\makeatletter\def\f@size{8.5}\check@mathfonts
\begin{align*}
\text{SOP}_{n}=1-F_{ \gamma _{e}^{n,\left(  T\right)} } \left(\frac {\theta -\zeta}{C_{n}^{g}}  \right)+\int\limits_{0}^{\frac{\theta -\zeta}{C_{n}^{g}}}  f_{ \gamma _{e}^{n,\left(  T\right)}} \left(y \right) F_{ \gamma _{n}^{n,\left(  T\right)}} \left( \zeta + C_{n}^{g} y  \right) \mathrm{d}y.
\end{align*}
\endgroup
\section{Proof Of Lemma \ref{lemma : gamma e}}
\label{appendix: gamma e}
We calculate $F_{\gamma _{e}^{n,\left(2 \right)}}\left(x\right)$ by following a similar  way as \cite{liu2017enhancing}. So we have:
\begin{align*}
F_{\gamma _{e}^{n,\left(2\right)}} \left(x \right) &= \mathbb{E} _{\Phi _{e}}    \left\lbrace \prod _{e \in \Phi _{e}, d_{m,e} \geq 0}  F_{ |{g}_{m,e}|^{2}}  \left( \frac{x \left( 1+d_{m,e}^{\alpha}\right)}{P_{C} }\right) \right\rbrace =
 \exp \left\lbrace   -\lambda_{e} \int\limits_{R^{2}} \left(  1-F_{|g_{m,e}|^{2}} \left(   \frac{x \left(1+ d_{m,e}^{\alpha}\right)}{P_{C} }  \right)   \right) r \, \mathrm{d}r \right\rbrace  \\
&= \exp \left\lbrace  -2\pi \lambda_{e} \int\limits _{0}^{\infty} r e^{-\frac{x \left( 1+r^{\alpha} \right)}{P_{C}}} \mathrm{d}r  \right\rbrace =
\exp \left[ -\chi_{1} \frac{e^{-\frac{x}{P_{C}}}\Gamma \left(  \eta   \right)}{x^{\eta}} \right],
\end{align*}
therefore $f_{\gamma _{e}^{n,\left( 2  \right)}} \left(  x \right)$ equals to:
\begin{align*}
f_{\gamma _{e}^{n,\left(2\right)}} \left(  x \right)=
\chi _{1} \Gamma \left( \eta \right) \exp \left[ -\chi_{1} \frac{e^{-\frac{x}{P_{C}}}\Gamma \left(  \eta   \right)}{x^{\eta}} \right] e^{-\frac{x}{P_{C}}}\left( \frac{\eta}{x^{\eta +1}} + \frac{1}{P_{C}x^{\eta}}\right).
\end{align*}
\section{Proof Of Lemma \ref{lemma: gamma eJ}}
\label{appendix: gamma eJ}
We derive $F_{\gamma _{e}^{n,\left( 2J \right)}} \left(x \right)$ and $f_{\gamma _{e}^{n,\left( 2J \right)}} \left(x \right)$ by following a similar approach as \cite{liu2017enhancing}:
\begingroup\makeatletter\def\f@size{9}\check@mathfonts
\begin{equation*}
\begin{aligned}
F_{ \gamma _{e}^{n,\left( 2J \right)}}\left( x \right)&=\mathbb{E}_{\Phi _{e}} \left\lbrace  \prod_{e \in \Phi_{e}, d_{m,e}\geq 0 }  F_{|{g}_{m,e}|^{2}}\left( \frac{x  \left( 1+d_{m,e}^{\alpha}\right)}{\beta P_{C} -\left(1-\beta  \right)P_{C}\,x} \right) \right\rbrace = 
\exp \left[ -\lambda_{e} \int\limits_{R^{2}}  \left( 1-  F_{|{g}_{m,e}|^{2}}\left( \frac{x   \left( 1+d_{m,e}^{\alpha}\right)}{\beta P_{C} -\left(1-\beta  \right)P_{C}\,x} \right)  \right) r \, \mathrm{d}r \right]\\
&=\exp \left[  -2\pi \lambda_{e}\int\limits_{0}^{\infty} r e^{-\frac{x\left(1+ r^{\alpha}\right)}{\beta P_{C} -\left(1-\beta  \right)P_{C}\,x}} \mathrm{d}r \right] U \left( -x+\frac{\beta}{1-\beta} \right) +U \left(  x-\frac{\beta}{1-\beta} \right)=U \left(  x-\frac{\beta}{1-\beta} \right)\\
&+ \exp \left[  -\chi_{2} e^{-\frac{x}{\beta P_{C} -\left(1-\beta  \right)P_{C}\,x}} \left(  \frac{\beta P_{C} -\left(1-\beta  \right)P_{C}\,x}{x}  \right) ^{\eta} \, \right] U \left( -x+\frac{\beta}{1-\beta} \right),
\end{aligned}
\end{equation*}
\endgroup
so we have:
\begingroup\makeatletter\def\f@size{9}\check@mathfonts
\begin{equation*}
\begin{aligned}
&f_{\gamma _{e}^{n,\left( 2J \right)}}\left( x \right)=\chi_{2}\beta P_{C}e^{-\frac{x}{ \beta P_{C}-\left( 1-\beta \right) P_{C}x}}
  \exp \left[  -\chi_{2} e^{-\frac{x}{\beta P_{C} -\left(1-\beta  \right)P_{C}x}} \left(  \frac{\beta P_{C} -\left(1-\beta  \right)P_{C}\,x}{x}  \right) ^{\eta} \, \right]  \\
&\times \left(\frac{\left( \beta P_{C} -\left(1-\beta  \right)P_{C}x  \right)^{\eta -2}}{x^{\eta}}+\eta \frac{\left( \beta P_{C} -\left(1-\beta  \right)P_{C}x  \right)^{\eta -1}}{x^{\eta +1}}\right) U \left(  -x+\frac{\beta}{1-\beta}\right).
\end{aligned}
\end{equation*}
\endgroup

\end{document}